\documentstyle[aps,prl,multicol,epsf]{revtex}

\begin{document}
\def\CC{{\rm\kern.24em \vrule width.04em height1.46ex depth-.07ex
\kern-.30em C}}
\def\P{{\rm I\kern-.25em P}}
\def\RR{{\rm
         \vrule width.04em height1.58ex depth-.0ex
         \kern-.04em R}}

\draft
\title{Topological protection and quantum noiseless subsystems}
\author{{ Paolo Zanardi$^1$ and Seth Lloyd$^2$ }}
\address{
$^1$  Institute for Scientific Interchange  (ISI) Foundation,
Viale Settimio Severo 65, I-10133 Torino, Italy\\
Istituto Nazionale per la Fisica della Materia (INFM)\\
$^2$ Department of Mechanical Engineering\\
Massachusetts Institute of Technology, Cambridge Massachusetts 02139 
}
\date{\today}
\maketitle
\begin{abstract}
Encoding and manipulation of quantum information by means of topological
degrees of freedom provides a promising way to achieve natural fault-tolerance
that is built-in at the physical level.
We show that this topological approach to quantum information
processing is a particular instance of the notion of computation in a noiseless
quantum subsystem. The latter then provide the most general conceptual
framework for stabilizing quantum information and for preserving quantum coherence
in topological and geometric systems.
\end{abstract}

\begin{multicols}{2}

Quantum information is sensitive: more bad things can happen to a quantum bit
than to a classical bit \cite{QC}.  A variety of schemes for protecting quantum information
have been developed, including quantum error correction codes \cite{ERR}, decoherence free subspaces
\cite{EAC}, noiseless subsystems \cite{KLV}, bang-bang decoupling \cite{SUPPR},
 and topological quantum computation \cite{TOP}.
The first four of these techniques are closely related to each other and can be described
in a simple unified framework based on representations of the algebra of errors \cite{KLV,stab}.
 This paper shows that  topological quantum computation also falls into the error-algebra framework.
This result suggests that methods for preserving quantum coherence in general
fall within a unified algebraic framework.

In  the error-algebra framework quantum information is protected by using symmetry.
The symmetry that protects quantum information can either exist naturally in
the interaction of the quantum information processing system with its environment,
as in the case of decoherence free subspaces and noiseless subsystems; or the
symmetry can be induced by adding additional dynamics as in the case of bang bang
decoupling; or the symmetry can exist implicitly as in quantum error correcting codes.
The role of explicit, dynamical, and implicit symmetries in stabilizing quantum states
and preserving quantum coherence of course goes far beyond quantum information processing:
preservation of coherence via symmetry plays a role in virtually all quantum systems.
One apparent exception to this rule is the case of topological quantum systems, in
which topological degrees of freedom are intrinsically resilient to local errors.
Although topological quantum computation is related to toric error correcting codes \cite{toric},
the physical mechanism by which it preserves quantum information goes
beyond toric codes.  It is an interesting question, then, whether topological quantum
computation in particular, and topological quantum systems in general,
can be treated in a unified framework along with the above mechanisms.
This paper shows that the answer to this question is Yes.

First, review briefly the way in which symmetry protects quantum information \cite{KLV,stab}.
Suppose that one has a quantum system $S$ with Hilbert space ${\cal H} \cong {\CC}^{d} $, interacting
with an environment.  The effect of the environment on the system is given by
a set of error operators $\{E_\alpha\}$: each $E_\alpha$ represents some bad thing that can
happen to the quantum system. Sums of arbitrary products of error operators together with their Hermitian
conjugates generate the error algebra $\cal A$.
This error algebra is the fundamental object in the algebraic approach to protecting
quantum information: it contains
all the information about the quantum information-stabilizing strategies.
Let ${\cal A}' = \{ X: [X,E_\alpha]=0\}$ be the commutant of the error algebra.
The elements of the unitary part ${\cal U}( {\cal A}')$ of ${\cal A}^\prime$ are
symmetries for the error algebra ${\cal A}.$

The degrees of freedom associated with observables in ${\cal A}^\prime$
are {\em noiseless} ones \cite{KLV,stab}: they are by definition decoupled from
the noise processes enacted by the elements of $\cal A.$    These noiseless
observables give rise to decoherence free subspaces and noiseless subsystems as follows.
It is a basic theorem of representations of algebras that
the Hilbert space $\cal H$ then decomposes as follows
\begin{equation}
{\cal H}\cong \oplus_{J} \CC^{n_J}\otimes \CC^{d_J}.
\label{split}
\end{equation}
where the $J$ label the different irreducible representations of the algebras $\cal A$ and ${\cal A}'$,
$d_J$ is the dimension of the $J$'th irreducible representation of ${\cal A}$, and $n_J$
is the dimension of the $J$'th irreducible representation of ${\cal A}'$.
(Formally, this decomposition
of the Hilbert space into sums of tensor product spaces corresponds to the
so-called central decomposition \cite{ALG}:
${\cal A}\cong\oplus_{J\in{\cal J}} \openone_{n_J}\otimes M(d_J,\,\CC)$,
where $M$ is the set of $d_J \times d_J$ matrices over $\CC$.)
The tensor product
structure arises naturally because members of $\cal A$ and ${\cal A}'$ commute: in each
term in the sum, the error operators in $\cal A$ act on the subsystem $\CC^{d_J}$ while
leaving the {\em noiseless subsystem} $\CC^{n_J}$ unchanged.  The decomposition (\ref{split})
shows that nontrivial noiseless subsystems exist only when ${\cal A}$ has a
noncommutative symmetry group ${\cal G} \equiv {\cal U}({\cal A})$.
In the particular case in which $d_J=1$ one has an instance of a noiseless code or
decoherence-free subspace \cite{EAC}.  Bang-bang decoupling is a method for inducing
an effective symmetry in the error dynamics that gives rise to effective noiseless
subsystems.  Finally, in \cite{KLV}, it was shown that
this tensor product decomposition is at the root of quantum error correcting codes:
errors act on the subsystem $\CC^{d_J}$ while
the quantum information lying in the encoded subsystem $\CC^{n_J}$ remains unchanged.
So virtually all known methods for protecting quantum information fall within
the error-algebra formalism.  

This formalism is also at the root of performing
quantum information processing in a fault-tolerant fashion.
Quantum manipulations within a noiseless
subsystem can be performed by applying transformations
from ${\cal A}^\prime.$  This last technique allows one to perform universal quantum computation
using quantum logic gates (such as swap gates) that are not universal on the entire
Hilbert space, a phenomenon known as encoded universality \cite{enc}.

For what follows it is important to notice that the state-space structure (\ref{split})
is reminiscent of {\em superselection} \cite{giulini}. In superselection theory
the algebra $\cal A$ is viewed as the one generated by the {\em whole} set of physical observables
rather than the one associated to  a set of distinguished interactions (the error operators).
In this context
${\cal U}({\cal A'})$ is called the gauge group.
The operators of $\cal A$ are not able to change the quantum numbers associated
with the gauge transformations and the state space accordingly splits in a direct sum
of non-connected sectors.
Accordingly the elements of the gauge group are operators that
commute with {\em all} the physical quantities and their eigenvalues therefore cannot
be changed by any physical operation. It is well-known that such a situation can occur
only in the cases in which ${\cal A}$ describes an infinite set of degrees of freedom,
the paradigmatic case being provided by field theory \cite{haag}.
The different sectors  describe now different inequivalent {\em phases} in which the system
can exist; a major illustration of this state of affairs is provided by the phenomenon of
spontaneous symmetry breaking.

Another very important occurrence of superselection is given by the quantization
of systems whose classical configuration manifold $\cal M$ has non-trivial
topology, e.g., with fundamental group $\pi_1({\cal M})\neq Id$ \cite{NAK}.
In this case the superselection sectors correspond to inequivalent quantizations
labelled by irreducible representations of $\pi_1({\cal M})$ \cite{mora}.
When $\cal M$ is the manifold associated with $N$ indistinguishable particles
living in $d$ dimensions the different irreducible representations
describe different quantum {\em statistics} \cite{any}.
For $d\ge 3,$  $\pi_1({\cal M})$ is given by the permutation group ${\cal S}_N$,
while for $d=2$ the fundamental group is the braid group ${\cal B}_N$. Particles associated with
one-dimensional irreducible representations of ${\cal B}_N$ are called abelian
{\em anyons}, and particles associated with higher dimensional representations are
called nonabelian anyons.
This latter class of systems is exactly the one that has been argued to be useful
for quantum computation \cite{TOP}.

Now we apply the error algebra framework to topological information protection.
The prototype system in which we are interested is a lattice $\Lambda$
having attached to each of its sites (or edges) a finite-dimensional
quantum system, e.g., a qubit, with state space ${\cal H}_i$.
The lattice $\Lambda$ is supposed to be embeddable
in a  two-dimensional surface $\cal M$ with {\em genus} $g.$
For example, the lattice could be a square lattice with periodic
boundary conditions, embedded in a torus.  The interaction among the
local quantum systems is described by a {\em local} Hamiltonian
$H_\Lambda$ having a $D(g)$-dimensional degenerate ground state
${\cal C}\subset {\cal H}_\Lambda\equiv\otimes_{i\in\Lambda} {\cal H}_i$,
where $D$ is an exponential function of its argument.
${\cal C}$ is the {\em code} subspace.
The key idea is that $H_\Lambda$ is designed
in such a way that the  ground-state degeneracy has a topological origin.
This means that orthogonal elements of $\cal C$ have to
correspond to different eigenvalues of global observables.

For some of the systems studied in \cite{TOP,ioff}
a complete set of commuting global observables in $\cal C$ can be constructed as follows.
Let  $\{\gamma\}_{i=1}^{2g}$ denote the set of non-contractible loops generating the homology
group $H_1({\cal M})$ of $\cal M$ \cite{NAK}.
 One can consider the operators $X_\gamma\equiv\prod_{i\in\gamma} x_i$
where the $x_i\in\mbox{End}({\cal H}_i)$ are suitable site  operators e.g., $\sigma^z_i.$
For the sake of simplicity assume that the $X_\gamma$'s are hermitian self-inverse
operators i.e., $X_\gamma^2=\openone.$
If this is the case $\cal C$ can be decomposed in terms of the $2^{2g}$
joint eigenvectors of the
$X_{\gamma_i}$ i.e., ${\cal C}=\mbox{span}\{|J\rangle\equiv|j_1,\ldots,j_{2g}\rangle\}$ where
$X_{\gamma_i}\,|J\rangle=j_i |J\rangle
\, (j_i\in{\bf Z}_2\equiv\{-1,1\}).
$

We denote by ${\cal A}_{glob}$ the abelian algebra generated
by the $X_{\gamma_i}.$
Local operators $X\in{\cal A}_{loc}\equiv{\cal A}_{glob}^\prime$ cannot by definition
modify the global properties described by the $X_{\gamma_i}$'s.  Local operators
therefore a) cannot induce tunneling between orthogonal ground states, and
b) cannot distinguish elements in the code subspace $\cal C.$
It follows that $\forall X \in {\cal A}_{loc}$ one has
$ \langle J^\prime|\,X\,|J\rangle = \delta_{J,J^\prime} c(X),$
where $c\colon {\cal A}_{loc}\mapsto \CC.$
If $\Pi_{\cal C}$ is the projector over the ground-state
$\cal C$ a compact way to express the condition above is given by
\begin{equation}
\Pi_{\cal C} \,X\,\Pi_{\cal C} =c(X)\,\Pi_{\cal C},\quad\forall X\in{\cal A}_{loc}
\label{kl}
\end{equation}
This latter relation amounts to saying that $\cal C$ behaves as an error correcting code
with respect the class of errors represented by local operators \cite{ERR}.

The  vector space  generated by
the action of local operators over $\cal C$ comprises the whole of ${\cal H}_\lambda$.
Relations (\ref{kl}) above imply that  the subspaces ${\cal A}_{loc}|J\rangle$
for different $J$'s are orthogonal and {\em isomorphic}.
It follows that
one has
the following splitting according to the
irreducible representations of ${\cal A}_{glob}$: 
\begin{equation}
{\cal H}_\Lambda=\oplus_{J\in{\bf Z}_2^{2g}} {\cal A}_{loc}|J\rangle
\cong{\cal C}^\prime\otimes{\cal C}
\label{simple-split}
\end{equation}
where ${\cal C}^\prime$ is a $2^{\ln{\mbox{dim}{\cal H}_\Lambda}-2g}$-dimensional
factor associated with {\em local} degrees of fredom.
This factor is associated with the syndrome measurements
for quantum error correction, its local  nature implies that those
measurements can be performed with elements belonging to ${\cal A}_{loc}.$

By comparing Eq. (\ref{simple-split})  with Eq. (\ref{split})
is not difficult to realize  that  the following holds:

\smallskip
{\em Proposition.}
The topologically protected sector $\cal C$ of ${\cal H}_\Lambda$ can be identified with
a noiseless subsystem with respect the error algebra generated by local interactions.
The associated gauge group is generated by operators
with non-trivial topological content.

\smallskip
It is quite important to make clear  that all these properties are meant to hold in the
limit in which the size $|\Lambda|$ goes to infinity;  for finite size systems they are
fulfilled only in an approximate way, though with exponential accuracy \cite{TOP}.
Formally equation (\ref{kl}) has then  the meaning:
\begin{equation}
\|\Pi_{\cal C} \,X\,\Pi_{\cal C} - c(X)\,\Pi_{\cal C}\|= O(e^{-\alpha |\Lambda|^{1/n}})
\end{equation}
where $\alpha>0$ and $n$ is an integer.
It is just when $|\Lambda|\mapsto\infty$ the ${\cal A}_{loc}|J\rangle$ become truly disjointed
sectors. In this limit  one has an effective  state-space splitting
having a topological origin  also known as homotopical superselection rule \cite{mora}.
In  the finite $|\Lambda|$ case there is always a (small) chance of a local perturbation
inducing  tunnelling between different $|J\rangle$'s and  of having
different diagonal elements in ${\cal C}.$

The result stated in the proposition above --- in view of the general connections between error
correcting codes and noiseless subsystems already established in \cite{KLV} and \cite{stab} ---
is not conceptually
totally surprising. On the other hand it represents a novel and 
natural physics-based instance of those connections. Moreover
in this way we are also pointing out that, after tracing over the local degrees of freedom
i.e., ${\cal C}^\prime,$ one need not perform any active recovery from the error:
a completely passive stabilization is achieved.

A very interesting situation is when our system is gapped,
this means that there is a finite energy $\Delta$ between the ground and first excited states
that remains finite even in the large size limit.  In this case,
every local modification of the ground-state results in a finite increase of energy:
the system is {\em incompressible}. Prototypes of this kind of systems
are provided by fractional quantum Hall effect fluids and spin liquids \cite{frad}.

In this case  small  perturbations $X\in{\cal A}_{loc}$ are ineffective
in inducing tunnelling between the ground and the excited states (not just among ground states).
Indeed from elementary perturbation theory one has that the amplitude for
those  tunneling process scale as $k/\Delta<<1,$
where $k$ is the typical strength of matrix element of $X$ between ground and excited states.
Reasoning again in perturbation-theoretic fashion it is clear
that this relation along with Eq. (\ref{kl}) implies that ground state degeneracy
is {\em robust} against small and local perturbations.
 This is a signature of its topological
nature. It has been recently argued that this kind of stability result might be generic
for  quantum spin systems in lattices with short range interactions \cite{ragi}.
Moreover some (exotic) spin models whose ground states are robust for all
weak enough local pertubations have been explicitly constructed \cite{freed}

{\em Quantum information manipulation.}
The primary purpose of this paper is to identify in detail the connection
between topological protection of quantum information and the error-algebra
formalism of protecting quantum information via noiseless subsystems, quantum
error correcting codes, etc.  Now that that task has been performed, 
we would like to use the general formalism developed above to address the important problem
of the manipulation of topologically encoded quantum information.
The actual way in which universal topological quantum computation is performed
depends strongly on the underlying physical models\cite{TOP,ioff}:
the error-algebra formalism allows us to abstract certain common features of
these models.

Within the described  error-algebra framework, it is possible to describe how to
perform quantum computation within the code subspace by creating local excitations
and by moving them around the lattice.
The key point here is that in the topological models we are examining
the Hamiltonian spectrum comprising (non-abelian) localized  anyonic excitations \cite{TOP}.
By spatially exchanging those excitations one can enact
 operations  that are  able to induce coupling between  different topological sectors.
Moreover these operations will depend just on some global i.e., topological, feature
 of the exchange and therefore are stable against any local perturbation.

Acting on ${\cal C}$ with $N$-site operators $X^{\alpha_k}_{j_k}$ creates
an excited state with $N$ local excitations (the $\alpha$'s labels the different possible
``colors ''). These excited states are
degenerate as long as the the $j$'s are kept far apart \cite{TOP}.
One can then build a degenerate subspace endowed with a bi-partite (local and global)
tensor product structure
${\cal H}^N_\alpha:=\mbox{span}\{\prod_{k=1}^N X^{\alpha_k}_{j_k}\,{\cal C}\,/\,
l\neq k\Rightarrow j_\l\neq j_k\}\cong {\cal C}^\prime_\alpha \otimes {\cal C} .
$
The basis states in ${\cal H}_J^N(\alpha)$ are labelled by the locations $j_1,\ldots,j_N$
of the $N$ local excitations e.g., anyon-antianyon pairs.
In order to perform quantum manipulations one resorts to the anyonic nature
of these excited states.  Excitations can be moved about the lattice, either by
applying local dynamical swap operations (\cite{LLO}), or by dragging them
adiabatically e.g., using an external potential \cite{anyBCZ}, along some path
with non-trivial braiding pattern $b$.
Moving one excitation around another enacts an element of the braid group ${\cal B}_N$,
which in turn performs a quantum logic operation on the quantum information registered
in the code subspace. Finally the excitations are annihilated (fusion): the result of
the computation is registered in the local state of the system after fusion
has taken place.   

We denote by $\rho$ the particular high-dimensional representation of the Braid
group model involved in the given topological model.
The sequence of excitation, braiding, and fusion can be
 schematically summarized by the following map chain
\begin{equation}
{\cal C}\stackrel{\prod_k X_k}{\longrightarrow} {\cal H}^N_\alpha\stackrel{\rho(b)}
{\longrightarrow}
{\cal H}^N_\alpha \stackrel{\prod_k X^\dagger_k}{\longrightarrow}{\cal C},
\label{chain}
\end{equation}
it is important to note here that only the intermediate braiding step $\rho(\gamma)$
has a nontrivial topological content
and can therefore change the global quantum numbers.

The braid elements $\rho(\gamma)$ are expressible as an holonomy of a suitable
statistical connection \cite{any}; from this perspective the computational scheme
sketched above provides a particular instance of the so-called holonomic approach
to quantum computation \cite{HQC}.
In this kind of scheme information is encoded in a degenerate eigenspace
of a parametric family of (isodegenerate) Hamiltonians and manipulated
by driving the parameters along suitable adiabatic paths.
This enact a transformation of the encoding space into itself
via the holonomy associated with the  Wilczek-Zee non-abelian connection
i.e., gauge potential, generated by the Hamiltionian family \cite{WIZE}.
In the topological  case under examination   the  manifold of control parameters is given
by the  the set of  the coordinates of the anyonic excitations themselves.
When the statistical connection  has
an holonomy group coinciding with whole set of unitary transformations
over $\cal C$ the full computational power is achieved \cite{HQC}.
In this case universal fault-tolerant manipulations can be performed
on the coding ground state ${\cal C}.$  (Such computation is an example of
encoded universality.)  The common holonomic nature of geometric and topological
quantum computation suggests that conceptually there is a sort of continuous
path from purely geometric to purely topological quantum information processing schemes.
In order to optimize the fault-tolerance features one might think of designing
non-abelian Wilczek-Zee connections  with maximal topological content.

{\em Conclusions.}
In this paper we discussed the relation between the topological approach
to fault-tolerant quantum information processing and the quantum
error correction-avoidance  strategies. A unified view of this latter class
of by-now standard techniques is provided by the algebraic notion
of noiseless subsystem. We showed that this notion is powerful enough
to encompass even the former class: topologically protected
quantum codes are an instance of noiseless subsystem.
The crucial point consists  in the separation of local and global degrees of freedom by
means of the associated observable algebras. Morever we pointed out how
information  processing  within this kind of noiseless subsystems is then achieved
through the holonomic  manipulations of (non-abelian) anyonic excitations.


\end{multicols}
\end{document}